\documentclass[fleqn,usenatbib,letter,lineno]{mnras}
\usepackage{newtxtext,newtxmath}
\usepackage[T1]{fontenc}

\DeclareRobustCommand{\VAN}[3]{#2}
\let\VANthebibliography\thebibliography
\def\thebibliography{\DeclareRobustCommand{\VAN}[3]{##3}\VANthebibliography}

\usepackage{graphicx}	
\usepackage{amsmath}	

\title[TDEs are kickin']{What's kickin' in partial tidal disruption events?}

\author[Coughlin \& Nixon]{Eric.~R.~Coughlin,$^{1}$\thanks{ecoughli@syr.edu}
and C.~J.~Nixon$^{2}$
\\
$^{1}$Department of Physics, Syracuse University, Syracuse, NY 13210, USA \\
$^{2}$School of Physics and Astronomy, Sir William Henry Bragg Building, Woodhouse Ln., University of Leeds, Leeds LS2 9JT, UK
}

\date{Accepted XXX. Received YYY; in original form ZZZ}

\pubyear{2025}

\begin{document}
\label{firstpage}
\pagerange{\pageref{firstpage}--\pageref{lastpage}}
\maketitle

\begin{abstract}
Stars partially destroyed by a supermassive black hole (SMBH) in a partial tidal disruption event (TDE) can be ejected from the SMBH. Previous investigations attributed this positive-energy/velocity kick to asymmetries in the mass lost by the star near pericenter. We propose that asymmetric mass loss is not predominantly responsible for ``kicking'' the star, and that these kicks instead arise from the combination of a) the reformation of the core following an initial phase of quasi-ballistic motion, and b) the differential shear between the unbound and marginally bound (to the SMBH) material during this phase. We predict that the kick speed $v_{\rm kick}$ is weakly dependent on the stellar properties, and for SMBH masses $M_{\bullet} \gtrsim 10^{3} M_{\odot}$, $v_{\rm kick}$ is independent of SMBH mass, is not limited to the stellar escape speed $v_{\rm esc}$, and is related to the surviving core mass $M_{\rm c}$ approximately as $v_{\rm kick} \simeq 0.45 \left(M_{\rm c}/M_{\star}\right)^{-1/3}$, where $M_{\star}$ is the original stellar mass. For $M_{\bullet} \lesssim 10^{3} M_{\odot}$, we find that the maximum-attainable kick speed depends on SMBH mass, satisfies $v_{\rm kick, max} \simeq 0.4 \, v_{\rm esc}\left(M_{\bullet}/M_{\star}\right)^{1/6}$, and is reached for core masses that satisfy $M_{\rm c}/M_{\star} \lesssim 1.7\left(M_{\bullet}/M_{\star}\right)^{-1/2}$. This model predicts that massive stars with $M_{\star}\gtrsim few\times 10 M_{\odot}$ could be ejected at speeds $\gtrsim 1-2\times 10^3$ km s$^{-1}$ if stripped of $\gtrsim 50\%$ of their mass.
\end{abstract}


\section{Introduction}
\label{sec:intro}
Many tidal disruption events (TDEs) per year \citep{hills75, lacy82, rees88} are discovered by surveys such as ASAS-SN \citep{shappee14}, ZTF \citep{bellm19}, and -- imminently -- the Rubin Observatory/LSST \citep{ivezic19} (e.g., \citealt{vanvelzen21, hammerstein23, yao23, guolo24b}; see \cite{gezari21} for a review of TDE observations). A subclass of TDEs, known as repeating partial TDEs (rpTDEs), has also been discovered (e.g., \citealt{payne21, wevers23, liu23, evans23, guolo24, somalwar23, hinkle24, sun25, makrygianni25}); in these systems, the first accretion flare is followed by at least one other (comparable in brightness to the first) months to years later\footnote{Other systems have recently been discovered that display X-ray-only outbursts with recurrence times $\lesssim $ few days (e.g., \citealt{miniutti19, giustini20, song20, arcodia21, chakraborty21, wevers22, miniutti23, arcodia24, nicholl24, pasham24b}); these quasi-periodic eruptions could be related to partial disruptions \citep{king20}, but we do not consider these sources specifically here.}, as the star is stripped of only a fraction of its envelope during (at least the first) tidal encounter with the SMBH. In addition to providing strong evidence for the partial, as opposed to complete, disruption of a star by an SMBH, rpTDEs can uniquely probe the orbits of individual stars in distant nuclei and the ``fallback time,'' being the time taken for the stripped stellar material to first return to the SMBH and accrete \citep{wevers23, pasham24}.

Partial TDEs -- repeating or not -- have been studied numerically for decades, at least as early as the work by \cite{khokhlov93}, and the tidal interaction of two fluid bodies has been studied for much longer. Linear tidal theory predicts that the internal modes of the fluid bodies are excited at the expense of the orbital energy (i.e., the Keplerian energy associated with the center of mass of the orbiting object; \citealt{fabian75, press77}). Therefore, a natural conclusion is that the surviving core in a partial TDE is increasingly bound to the SMBH as the encounter becomes increasingly disruptive, e.g., a star that loses 50\% of its mass should be more strongly bound to the SMBH than one that loses 10\%. 

That this is not the case, and that a partially destroyed star is eventually ejected on a positive-energy orbit rather than increasingly tightly bound, was shown by \cite{manukian13} (see \citealt{faber05} for the same result in the context of planets partially destroyed by their host stars, and \citealt{kremer22, kiroglu23, vynatheya24} in the context of stars encountering lower-mass black holes), and \cite{cufari23} demonstrated that the maximum binding energy produced through the tidal interaction of a low-mass star (i.e., modeled as gas-pressure-dominated and polytropic) with a SMBH is $\sim 2\%$ of the stellar binding energy. This global minimum in the energy of the captured-star orbit occurs (for a low-mass star modeled as a $5/3$-polytrope) when the pericenter distance of the star is $\sim 1.85\,r_{\rm t}$ ($r_{\rm t} = R_{\star}\left(M_{\bullet}/M_{\star}\right)^{1/3}$ is the canonical tidal radius, with $R_{\star}$ and $M_{\star}$ the stellar mass and radius and $M_{\bullet}$ the SMBH mass), or $\beta \simeq 0.54$, where $\beta$ is the ratio of the tidal radius to the pericenter distance. \cite{manukian13} proposed that the origin of this positive-energy kick is asymmetry in the mass lost by the star near pericenter, and this explanation has been reiterated in more recent works (e.g., \citealt{gafton15, chen24}). 

Here we propose an alternative mechanism for the origin of positive-energy kicks in partial TDEs, which is related to the fact that near-complete disruptions -- which also have the highest-energy kicks \citep{manukian13, gafton15, miles20, cufari23} -- are accompanied by the \emph{reformation} of a core well after pericenter \citep{guillochon13, miles20, nixon21}. In Section \ref{sec:physical} we describe this model and show that the predictions are in agreement with the results of all past works, and we summarize and conclude in Section \ref{sec:summary}.

\section{Physical model}
\label{sec:physical}
\subsection{The argument}
In the complete disruption of a star, the positive-energy tail tends toward homologous expansion, such that -- assuming the stream is balanced in the transverse direction by self-gravity and the material is gas-pressure dominated -- the density of an individual fluid element declines with time as $\propto t^{-3/2}$ \citep{coughlin16}. Conversely, the larger shear near the zero-energy orbit implies that the density of the marginally bound fluid element declines with time as $\propto t^{-2}$. Therefore, if the maximum in the density along the stream originally coincides with the zero-energy orbit, it will systematically shift toward positive energies owing to the less rapidly declining density in the unbound tail. If the core reforms at some later time near the maximum-density fluid element, it will therefore be unbound. This is the argument and the physical origin we propose: in disruptions that lead to a core with positive energy, there was some time post-pericenter during which the stellar debris expanded quasi-dynamically, after which a core reformed near the maximum-density fluid element. One can then approximate the final core energy by considering the ballistic motion of the maximum-density fluid element following a complete disruption, assuming that the intermediate phase (i.e., during which self-gravity along the stream is driving the recollapse, but before the core has definitively reformed) does not dramatically alter the final core energy. This latter assumption will be upheld to a good degree of accuracy, because self-gravity must overcome the shear itself -- not the difference in shear that is ultimately responsible for generating the non-zero core energy -- which is nearly constant in the vicinity of the marginally bound fluid element (i.e., the high degree of symmetry near the marginally bound radius, and hence effectively the maximum-density fluid element at early times, implies that self-gravity will decelerate fluid elements in the unbound tail at the same rate at which it accelerates fluid elements in the bound tail). We pursue this line of analysis in the next two sections.

\subsection{1D Approximation}
\label{sec:1D}
More quantitatively, assume the debris moves only in the radial direction (i.e., away from the SMBH) and ignore self-gravity and pressure in this direction\footnote{Self-gravity and pressure remain important in the transverse directions even in the limit of complete disruption; e.g., \citet{kochanek94}.}, such that the equation of motion for a radial orbit in the gravitational field of a point mass $M_{\bullet}$ is
\begin{equation}
    \frac{d^2r}{dt^2} = -\frac{GM_{\bullet}}{r^2}, \label{req}
\end{equation}
where $d/dt$ is the Lagrangian time derivative (i.e., for a given $r_0$, where $r_0$ is the initial position of a fluid element). We then make the change of variables $r\rightarrow \xi = r/R(t)$ and $t \rightarrow \tau = \ln (R(t)/R_{\rm i})$, where $R(t)$ is the marginally bound Keplerian radius that satisfies
\begin{equation}
    \frac{dR}{dt} = \sqrt{\frac{2GM_{\bullet}}{R(t)}}, \,\,\, R(0) = R_{\rm i},
\end{equation}
such that Equation \eqref{req} becomes
\begin{equation}
      2\frac{d^2\xi}{d \tau^2}+\frac{d\xi}{d \tau}-\xi = -\frac{1}{\xi^2}. \label{xieq}
\end{equation}
Taylor expanding the solution for $\xi$ about the marginally bound radius, 
\begin{equation}
    \xi(\tau,\xi_0) = 1+A(\tau)\left(\xi_0-1\right)+B(\tau)\left(\xi_0-1\right)^2+\mathcal{O}\left[\left(\xi_0-1\right)^3\right], \label{xisolgen}
\end{equation}
inserting the expansion into Equation \eqref{xieq} and equating like powers in $\left(\xi_0-1\right)$ gives
\begin{equation}
    \mathcal{L}\left[A\right] = 0, \,\,\, \mathcal{L}\left[B\right] = -3A^2, \,\,\, \mathcal{L} = 2\frac{\partial^2}{\partial\tau^2}+\frac{\partial}{\partial \tau}-3. \label{ABeqs}
\end{equation}
The general solution for $A(\tau)$ is
\begin{equation}
    A(\tau) = \left(1-c_1\right)e^{-3\tau/2}+c_1e^{\tau}, \label{Aofc1}
\end{equation}
where further specification of $c_1$ requires knowledge of the initial velocity profile of the debris. As an example, the frozen-in approximation asserts that the velocity of every fluid element at $\tau = 0$ is equal to that of the marginally bound orbit, i.e.,
\begin{equation}
    v(\tau = 0) = \frac{dr}{dt}(\tau = 0) = \sqrt{\frac{2GM_{\bullet}}{R_{\rm i}}},
\end{equation}
which in terms of the variable $\xi$ gives
\begin{equation}
    \frac{d\xi}{d\tau}(\tau = 0)+\xi(\tau = 0) = 1.
\end{equation}
Inserting our solution for $\xi(\tau)$ (Equation \ref{xisolgen}) in terms of $A$ (Equation \ref{Aofc1}) then yields $c_1 = 1/5$. Different initial conditions will of course lead to differnet values of $c_1$, but in general we require $c_1 > 0$ to ensure that fluid elements do not cross.\footnote{In principle $c_1$ could be negative, which would bind the core to the SMBH as the gas pressure simultaneously increases, and this could plausibly be achieved in situations where the self-gravitational field is significantly enhanced during compression; this would be consistent with the findings in \citet{nixon22}, who find that a core can reform on a bound orbit in deep TDEs.} The general solution for $B(\tau)$ can then be written down, but the term $\propto A^2$ in Equation \eqref{ABeqs} eventually dominates over the homogeneous terms, and hence 
\begin{equation}
    \xi(\tau,\xi_0) \simeq 1+c_1e^{\tau}\left(\xi_0-1\right)-\frac{3}{7}c_1^2e^{2\tau}\left(\xi_0-1\right)^2. \label{xiofxi0}
\end{equation}

The line density, being the density integrated over the cross-sectional area of the stream, is then (assuming that the width does not vary strongly with $\xi_0$) 
\begin{equation}
    \lambda(\tau,\xi_0) = \lambda_0(\xi_0)\left(\frac{\partial \xi}{\partial \xi_0}\right)^{-1}.
\end{equation}
If the initial density profile, $\lambda_0(\xi_0)$, is peaked at the zero-energy orbit, then 
\begin{equation}
    \lambda_0(\xi_0) \propto 1-\alpha\left(\frac{R_{\rm i}}{R_{\star}}\right)^2\left(\xi_0-1\right)^2
\end{equation}
to leading order in $\xi_0-1$, where $\alpha$ is a constant $\sim \left(\rho_{\rm c}/\rho_{\star}\right)^{2/3}$, where $\rho_{\rm c}$ ($\rho_{\star}$) is the central (average) stellar density, to ensure that the density declines on the scale of the stellar core radius, $R_{\rm c}$, which is approximately related to the stellar radius, $R_{\star}$, via $R_{\rm c} \simeq R_{\star}\left(\rho_{\rm c}/\rho_{\star}\right)^{-1/3}$ \citep{coughlin22b}. More specifically, if the density profile of the star is approximated as spherically symmetric at $R_{\rm i}$, then 
\begin{equation}
    \alpha = \frac{1}{2}\left(\int_0^{1}g(\eta)\,\eta\,d\eta\right)^{-1}, \label{alphagen}
\end{equation}
where $\eta$ is spherical radius measured from the center of the star in units of $R_{\star}$ and $g$ is the stellar density profile in units of the central density; $\alpha = 1$, 3.83, and 18.4 for a constant-density, $\gamma= 5/3$-polytropic, and $\gamma=4/3$-polytropic star, respectively, in rough agreement with $\alpha \sim \left(\rho_{\rm c}/\rho_{\star}\right)^{2/3}$. 

If $\partial\xi/\partial \xi_0$ were constant and the shear the same everywhere, then the location of the maximum density in the stream would remain at $\xi_0 = 1$ (i.e., $\partial\lambda/\partial\xi_0 = 0$ for $\xi_0 \equiv 1$). However, using Equation \eqref{xiofxi0} and maintaining the second-order term, we find that the maximum density in the stream occurs for 
\begin{equation}
    \xi_{\rm 0, c}(\tau) = 1+\frac{3c_1}{14\alpha}\left(\frac{R_{\rm i}}{R_{\star}}\right)^{-2}e^{\tau}, \label{xic0}
\end{equation}
which, since $c_1 > 0$, systematically shifts to unbound orbits; specifically, the Keplerian energy is
\begin{equation}
\epsilon = \frac{1}{2}v^2-\frac{GM_{\bullet}}{r},
\end{equation}
which, upon using $r = R\xi$ and
\begin{equation}
\begin{split}
    v = \frac{dr}{dt} = \frac{d}{dt}\left[R\xi\right] &= \sqrt{\frac{2GM_{\bullet}}{R}}\left(\frac{d\xi}{d\tau}+\xi\right) \\
    &= \sqrt{\frac{2GM_{\bullet}}{R}}\left\{1+\left(A+\frac{dA}{d\tau}\right)\left(\xi_0-1\right)\right\},
\end{split} 
\end{equation}
where the final equality holds to leading order in $\xi_0-1$, becomes
\begin{equation}
\begin{split}
    \epsilon = \frac{GM_{\bullet}}{R}\left(3A+2\frac{dA}{d\tau}\right)\left(\xi_0-1\right) &= \frac{5 GM}{R}c_1e^{\tau}\left(\xi_0-1\right) \\ 
    &= \frac{5GM_{\bullet}}{R_{\rm i}}c_1\left(\xi_0-1\right).
\end{split}
\end{equation}
In the final equality we used the fact that $\tau = \ln\left(R/R_{\rm i}\right)$. Inserting into this expression Equation \eqref{xic0} above for $\xi_{\rm c, 0}$ then shows that the energy of the maximum-density fluid element is
\begin{equation}
    \epsilon(\xi_{0, c}) \equiv \epsilon_{\rm c} = \frac{5GM_{\bullet}}{R_{\rm i}}c_1\times \frac{3c_1}{14\alpha}\left(\frac{R_{\star}}{R_{\rm i}}\right)^2 e^{\tau} = \frac{15 c_1^2}{14\alpha}\frac{GM_{\bullet}R_{\star}^2}{R_{\rm i}^3}e^{\tau}.
\end{equation}
Letting $R_{\rm i} = r_{\rm t} = R_{\star}\left(M_{\bullet}/M_{\star}\right)^{1/3}$ and again using the fact that $e^{\tau} = R/R_{\rm i} \propto T^{2/3}$, where $T$ is time in units of $R_{\rm i}^{3/2}/\sqrt{GM_{\bullet}}$ (which is also $\sim$ the dynamical time of the star with $R_{\rm i} = r_{\rm t}$), this implies that the velocity at infinity -- the ``kick velocity'' -- is
\begin{equation}
    v_{\rm kick} \simeq v_{\rm esc}T^{1/3}.
\end{equation}
In the next section we relax the one-dimensional assumption and also maintain numerical factors, but this shows that the shift in energy is independent of $M_{\bullet}$, the kick velocity increases with time of core reformation as as $\propto T^{1/3}$, and is of the order the escape speed of the star. The normalizations of the core energy and kick velocity -- namely that they are comparable to the stellar binding energy and escape speed -- can also be immediately deduced from the fact that this effect comes in at the next-highest-order (than quadrupole) moment of the SMBH gravitational field \citep{metzger22b}, though the dependence on core reformation time (which correlates with core mass; see Section \ref{sec:core-mass}) obviously does not. 

\subsection{2D model}
The previous section let the motion of the debris be one-dimensional, which is inaccurate near pericenter. To incorporate the angular motion of the fluid, we construct the line density of the gas by integrating over the transverse direction of the stream, while maintaining that the motion along the stream is still ballistic. Specifically, ``along the stream'' is tangent to the maximum-density curve, which is the set of points that characterize the maximum density in the stream at a given distance from the SMBH, and ``transverse to the stream'' is within the plane and locally orthogonal to the maximum-density curve; see Figure 1 in \citet{coughlin23}. By construction the pressure and self-gravitational forces are zero in the transverse direction as one moves along this curve, and the forces arising from the non-inertial nature of the frame are generally small relative to the tidal terms \citep{coughlin23}.

We further assume that the maximum-density curve is aligned with the pericenter vector at pericenter, and the velocity is perpendicular to this vector with a magnitude equal to $\sqrt{2GM_{\bullet}/R_{\rm i}}$. This setup is effectively the frozen-in approximation \citep{lacy82, bicknell83, lodato09, stone13} but only applied to the maximum-density curve; we are also only interested in the motion near the marginally bound Keplerian orbit, where we expect it to be a much better approximation than as applied to its radial extremities. Then the equations of motion for each fluid element (i.e., Newton's second law $d^2x^{i}/dt^2 = -GMx^{i}/r^3$, where $x^{i} = \{x,y\}$ are the Cartesian coordinates of a fluid element and $r = \sqrt{x^2+y^2}$ is its distance from the black hole) along the maximum-density curve are
\begin{equation}
    \frac{\partial^2 X}{\partial T^2} = -\frac{X}{\left(X^2+Y^2\right)^{3/2}}, \quad \frac{\partial^2Y}{\partial T^2} = -\frac{Y}{\left(X^2+Y^2\right)^{3/2}},
\end{equation}
where $X = x/R_{\rm i}$, $Y = y/R_{\rm i}$, and
\begin{equation}
    T = \frac{\sqrt{GM_{\bullet}}t}{R_{\rm i}^{3/2}},
\end{equation}
and the initial conditions are
\begin{equation}
    X(0) = X_0, \,\,\, Y(0) = 0, \,\,\, \frac{\partial X}{\partial T}(0) = 0, \,\,\, \frac{\partial Y}{\partial T}(0) = \sqrt{2}.
\end{equation}
Mass conservation implies that the line density $\lambda$ at some time $T$ is related to the original line density $\lambda_0$ via $\lambda \,ds = \lambda_0\, ds_0$, where
\begin{equation}
    ds = \sqrt{dX^2+dY^2}
\end{equation}
is the differential arc length along the stream. Since $ds_0 = dX_0$, this therefore yields
\begin{equation}
    \lambda(T,X_0) \frac{\partial s}{\partial X_0} = \lambda_0(X_0), \label{lambdaofT}
\end{equation}
Note that $X_0 = 1+\mathcal{O}(R_{\star}/R_{\rm i})$ with $R_{\star}/R_{\rm i} \simeq R_{\star}/r_{\rm t} \ll 1$.

We are interested in the region of the fluid near the marginally bound Keplerian orbit, so we can let, as in the previous section, 
\begin{equation}
    \lambda_0(X_0) \propto 1-\alpha\left(\frac{R_{\rm i}}{R_{\star}}\right)^2\left(X_0-1\right)^2. \label{lambda0}
\end{equation}
Then the position of the maximum density is, again to leading order in $X_0 - 1$,
\begin{equation}
    X_{\rm 0, c}-1 = -\frac{1}{2\alpha}\left(\frac{R_{\star}}{R_{\rm i}}\right)^2\frac{\partial}{\partial X_0}\left[\ln\left(\frac{\partial s}{\partial X_0}\right)\right]\bigg{|}_{X_0 = 1}, \label{X0max2D}
\end{equation}
and thus the Keplerian energy associated with the maximum density is
\begin{equation}
    \epsilon_{\rm c} = \frac{GM_{\bullet}}{R_{\rm i}}\left(X_{\rm 0, c}-1\right) 
    = -\frac{GM_{\bullet} R_{\star}^2}{2\alpha R_{\rm i}^3}\frac{\partial}{\partial X_0}\left[\ln\left(\frac{\partial s}{\partial X_0}\right)\right]\bigg{|}_{X_0 = 1}.\label{epscapp}
\end{equation}
Setting $R_{\rm i} = r_{\rm t}/\beta$, this becomes
\begin{equation}
\frac{\epsilon_{\rm c}}{\epsilon_{\star}} = \kappa 
\times\left(-\frac{\partial}{\partial X_0}\ln\left(\frac{\partial s}{\partial X_0}\right)\bigg{|}_{X_0 = 1}\right),\label{epsilonc} 
\end{equation}
where $\kappa \equiv \beta^3/(2\alpha)$. As for the 1D model, this shows that the energy is of the order the binding energy of the original star (with $R_{\rm i} \simeq r_{\rm t}$). The fact that $\kappa \propto \beta^3$ seemingly suggests a strong scaling of the energy with $\beta$, but we note that 1) the notion of the destroyed stellar material following quasi-ballistic orbits is only valid for $\beta$ near the value for complete disruption, being $\sim 0.9$ ($\sim 2$) for a $5/3$-($4/3$)-polytrope \citep{guillochon13, mainetti17}, and 2) the value of $\alpha$, which is a measure of the density profile of the star at the time it passes through pericenter, will also possess some dependence on $\beta$. Using these values of $\beta$ and the $\alpha$ given below Equation \eqref{alphagen}, we find $\kappa \simeq 0.095$ for a $5/3$-polytrope, while $\kappa \simeq 0.22$ for a $4/3$-polytrope. The dependence of $\kappa$ on stellar structure can be approximated by noting that the $\beta$ at which the star is completely destroyed scales approximately with the central stellar density, $\rho_{\rm c}$, and the average stellar density, $\rho_{\star}$, as $\beta \propto \left(\rho_{\rm c}/\rho_{\star}\right)^{1/3}$, as found empirically from numerical simulations utilising a fitting formula by \citet{lawsmith20, ryu20}, and as demonstrated by \citet{coughlin22b} can be recovered analytically in the limit of a more accurate calculation involving the maximum gravitational field within the star. With $\alpha \propto (\rho_{\rm c}/\rho_{\star})^{2/3}$, this suggests that $\kappa \propto \left(\rho_{\rm c}/\rho_{\star}\right)^{1/3}$, where the proportionality constant is $\sim 0.1$ to match the value appropriate to the $5/3$ polytrope. Furthermore, main-sequence stars obey a mass-radius relationship approximately of the form $R_{\star} \propto M_{\star}$ (e.g., \citealt{demircan91}), $\epsilon_{\star}$ is constant, and hence there is not a strong dependence on stellar type. Finally, the kick velocity is
\begin{equation}
    \frac{v_{\rm kick}}{v_{\rm esc}} = \sqrt{\kappa}\times \sqrt{-\frac{\partial}{\partial X_0}\ln\left(\frac{\partial s}{\partial X_0}\right)\bigg{|}_{X_0 = 1}}. \label{vkickfun}
\end{equation}

\begin{figure}
    \includegraphics[width=0.48\textwidth]{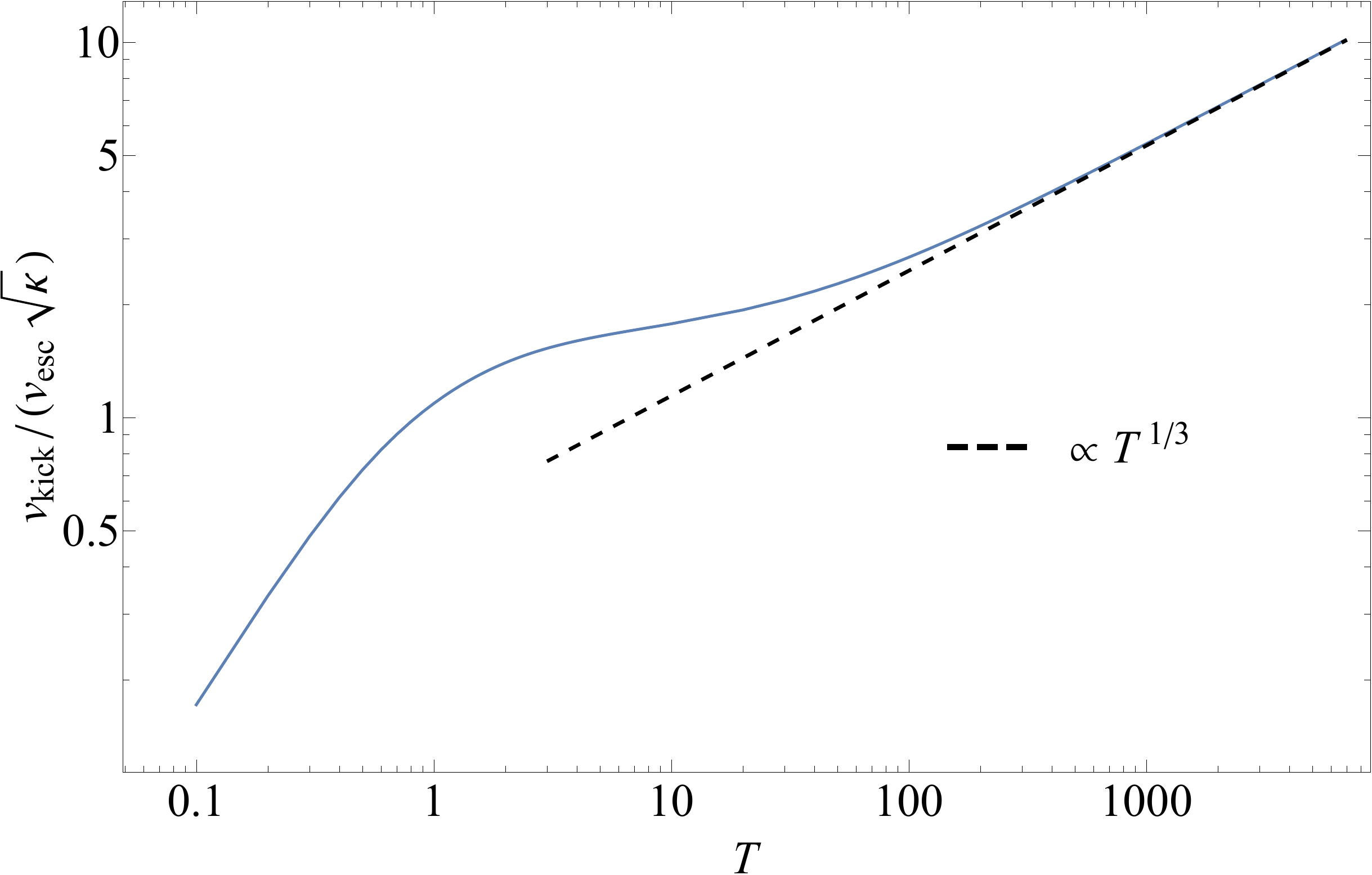}
    \caption{The kick velocity, normalized by $v_{\rm esc}\sqrt{\kappa}$, as a function of recollapse time $T$, where $T$ is in units of the dynamical time at pericenter (which is approximately the stellar dynamical time). The dashed line shows the $\propto T^{1/3}$ prediction from the 1D model.}
    \label{fig:X0max}
\end{figure}

Figure \ref{fig:X0max} shows the kick velocity normalized by $v_{\rm esc}\sqrt{\kappa}$ as a function of the recollapse time $T$ relative to the dynamical time at $R_{\rm i}$ (being $\sim$ the dynamical time of the star), implying that if a core reforms at time $T$, its velocity is given by $v_{\rm kick}(T)$. The black-dashed line shows the $\propto T^{1/3}$ scaling predicted from the one-dimensional model, demonstrating that (unsurprisingly) angular momentum is important at early times. After $\sim 50$ dynamical times the kick velocity approaches
\begin{equation}
    \frac{v_{\rm kick}}{v_{\rm esc}} \simeq 0.533\sqrt{\kappa} T^{1/3}, \label{vkickfun2}
\end{equation}
where the factor of $0.533$ results in the normalization of the black-dashed line in Figure \ref{fig:X0max}.

\subsection{The relation between core mass and core energy/kick-velocity}
\label{sec:core-mass}
Our argument is that the positive-energy kick imparted to the star in a partial TDE is determined by a phase of quasi-ballistic evolution, such that the difference in shear between the bound and unbound debris shifts the peak in the density to positive energies. The core is presumed to reform under the influence of self-gravity at some time $T$ post-disruption near the location of maximum density, at which point its energy is ``locked in'' at the value given by Equation \eqref{epsilonc}. The origin of the core reformation is almost certainly related to the fact that a gas-pressure-dominated, tidally disrupted debris stream is marginally self-gravitating, i.e., self-gravity in the directions transverse to the stream is just barely capable of counterbalancing the tidal shear along the stream. The debris stream is, in fact, overstable in the transverse directions, provided that the stream density is initially above a critical value; this critical value is always achieved in TDEs, and hence the oscillations grow in amplitude at a rate that is $\propto T^{1/6}$ and with a frequency that is exponential in time \citep{coughlin23}. 

The debris stream is therefore extremely weakly unstable, and the time at which the core reforms -- due to perturbations induced by the density profile of the original star and the coupling with the overstable oscillations in the transverse directions -- would be challenging to predict analytically with accuracy. Nonetheless, the mass of the core that reforms out of the stream should be approximately given by $\sim \rho H^3$, where $\rho$ and $H$ are the density and stream width at that time. The stream density near the marginally bound Keplerian radius declines with time as $\propto T^{-2}$, while the stream cross-sectional radius grows as $H \propto T^{1/3}$ (for a gas-pressure dominated stream; \citealt{coughlin23}), and hence the mass of the core -- at whatever time it forms -- is
\begin{equation}
    M_{\rm c} \simeq M_{\star} T^{-1}, \label{McofT}
\end{equation}
where we have neglected a numerical multiplicative factor that must be of the order unity (i.e., if the star recollapses near pericenter, then the core mass must be comparable to the original stellar mass) and that cannot be determined analytically (without a deeper understanding of the nature of the instability). Nonetheless, this numerical factor can be obtained empirically from hydrodynamical simulations; for example, by using the fitting function given in the caption of Figure 5 of \cite{gafton15} that relates the core mass to the kick velocity at any given value of the core mass, we can fix the numerical factor in Equation \eqref{McofT} to determine the kick velocity as a function of core mass. 

\begin{figure}
    \includegraphics[width=0.495\textwidth]{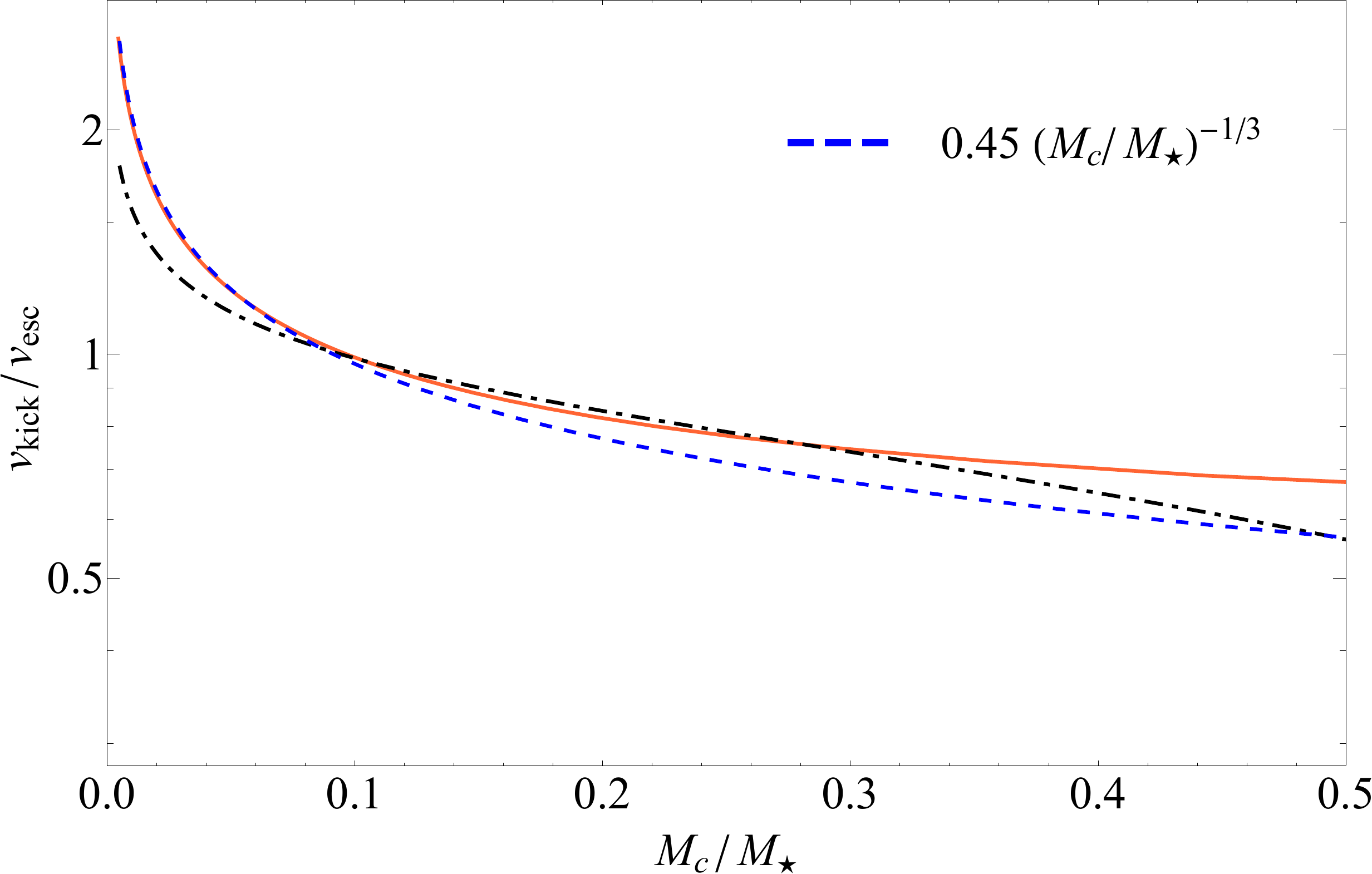}
    \caption{The kick velocity as a function of the core mass as determined from the model (orange curve), alongside the fitting function used in \citet{gafton15} (black-dashed), where the normalization is determined from the value of the fitting function at $M_{\rm core}/M_{\star} = 0.1$. The blue-dashed curve gives the approximate scaling from the model of $v_{\rm kick}\propto M_{\rm c}^{-1/3}$.}
    \label{fig:vkick}
\end{figure}

Figure \ref{fig:vkick} shows $v_{\rm kick}(M_{\rm c}/M_{\star})$ that results from using the value of their fitting function at $M_{\rm c}/M_{\star} = 0.1$ as well as the fitting function itself (black-dashed curve). The model prediction agrees well with their numerical results until $M_{\rm c}/M_{\star} \gtrsim 0.5$, where many of the approximations we made here likely break down; we also note that the differences between our model prediction and the fitted curve in \citet{gafton15} are within the $1-\sigma$ deviation between their fitted function and their simulation results (i.e., the gray region in their Figure 5). The behavior of $v_{\rm kick}$ at small core masses is $v_{\rm kick}\propto M_{\rm c}^{-1/3}$, which comes from the asymptotic scaling of $v_{\rm kick}\propto T^{1/3}$; including the normalization deduced from comparisons to the fitting function in \cite{gafton15} gives
\begin{equation}
    v_{\rm kick} \simeq 0.45\left(\frac{M_{\rm c}}{M_{\star}}\right)^{-1/3}. \label{vkickapp}
\end{equation}
Equation \eqref{vkickapp} suggests that the kick velocity is a weak function of the surviving core mass that, nonetheless, grows asymptotically as the core mass decreases. This curve is shown in Figure \ref{fig:vkick} (blue-dashed).

The breakdown between the agreement of this model and the results of hydrodynamical simulations is guaranteed once the core mass is a substantial fraction of the original stellar mass, which arises from the fact that we are assuming that self-gravity is completely ignorable for some finite time post-pericenter. Contrarily, the star only begins to lose mass for a pericenter distance comparable to $\sim 2 r_{\rm t}$, such that it is not valid (or at least much more questionable) to treat the gas as being in purely ballistic motion for any amount of time. Nevertheless, this physical origin could still be responsible for the stellar ejection at significantly larger core masses, but the influence of self-gravity would have to be included when considering the dynamics of the maximum-density fluid element. A more detailed analysis of, e.g., the velocity field in the vicinity of the maximum-density fluid element from hydrodynamical simulations would provide a means to test this hypothesis.

\subsection{Maximum core energy and kick velocity}
\begin{figure}
    \includegraphics[width=0.495\textwidth]{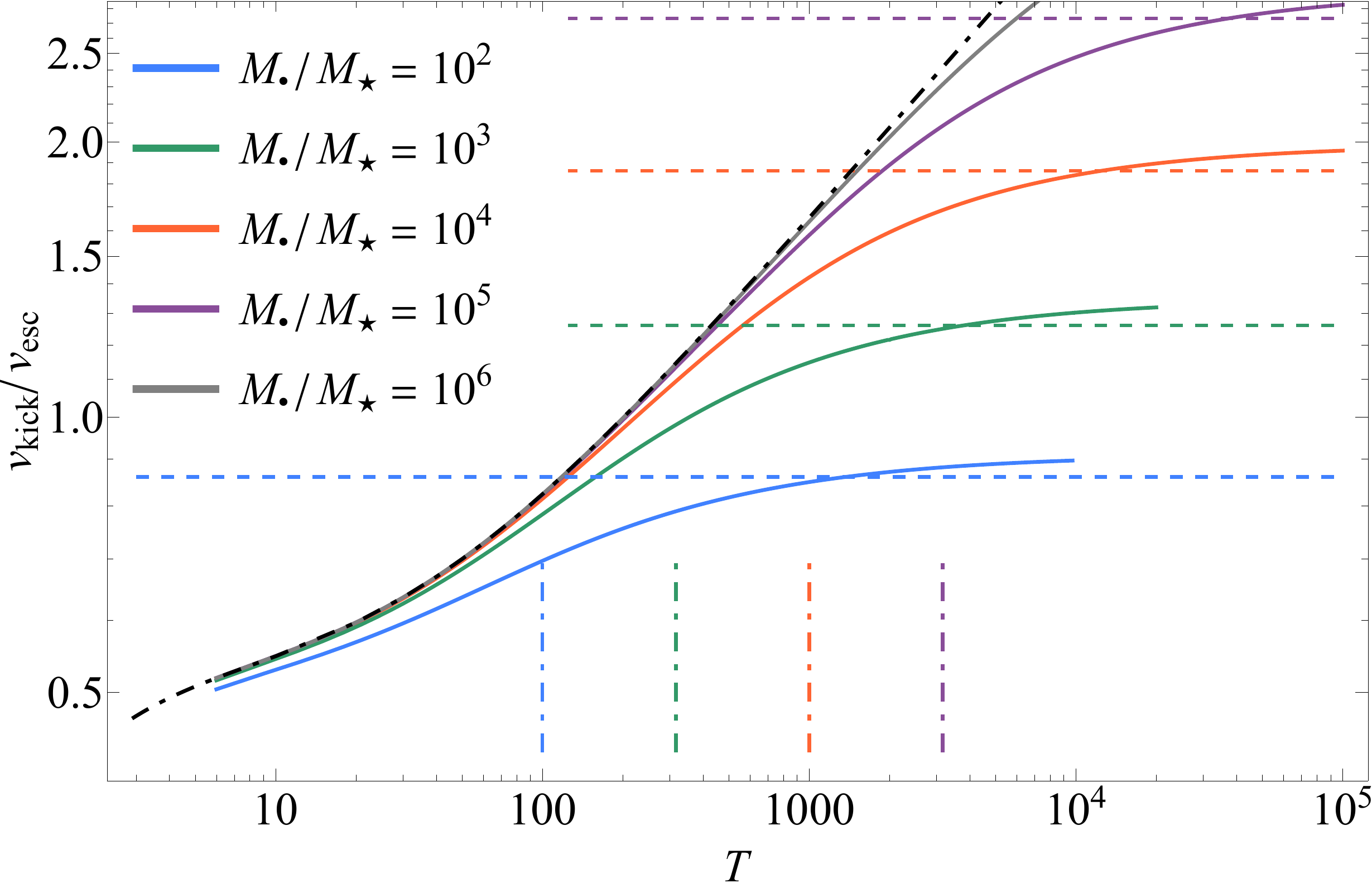}
    \caption{The numerically obtained kick velocity as a function of the time of core reformation $T$ (in $\sim$ units of the stellar dynamical time) for the SMBH mass ratios in the legend. The dashed lines are the approximate limits given by Equation \eqref{vmaxapp}, the vertical dot-dashed lines are the times at which these speeds are expected to be approached (Equation \ref{Tmax}), and the black dot-dashed curve is the $M_{\bullet}$-independent prediction (Equation \ref{vkickfun}), valid for $T \lesssim 10\left(M_{\bullet}/M_{\star}\right)^{1/2}$.} 
    \label{fig:vmax}
\end{figure}

The kick velocity from the one- and two-dimensional models asymptotically scales as $\propto T^{1/3}$, but this trend cannot continue indefinitely, because the density profile is preserved once the location of maximum density reaches homologous expansion, i.e., where the fluid elements have approximately reached their asymptotic velocities and the velocity profile is $\propto r$ (see Figure 2 of \citealt{coughlin16}). The time at which this occurs can be estimated by equating the velocity at the zero-energy orbit, $\sqrt{2GM_{\bullet}/R(T)}$, to the terminal velocity appropriate to energy $\epsilon_0$, $\sqrt{2\epsilon_0}$, which gives
\begin{equation}
      \left(\frac{3\sqrt{2}}{2}T\right)^{2/3} \simeq \frac{GM_{\bullet}}{\epsilon_0 R_{\rm i}}.
\end{equation}
Near this time the fluid element must transition to a constant velocity, as its speed cannot fall below the value guaranteed by energy conservation. Using this scaling in Equation \eqref{epsilonc} in the limit that the time-dependent term scales as $\sim T^{2/3}$ gives a maximum energy of\footnote{This energy must be less than $\epsilon_{\star}\left(M_{\bullet}/M_{\star}\right)^{1/3}$, i.e., the maximum energy spread imparted by the tidal field, which is guaranteed for any realistic stellar progenitor; see the discussion below Equation \eqref{epsilonc}.}
\begin{equation}
    \epsilon_{\rm max} =\epsilon_{\star}\times   0.4\sqrt{\kappa}\left(\frac{M_{\bullet}}{M_{\star}}\right)^{1/3},
\end{equation}
where the numerical factor incorporates the normalization of the $\propto T^{2/3}$ scaling, and a maximum kick velocity of
\begin{equation}
    v_{\rm kick, max} \simeq 0.4 v_{\rm esc}\left(\frac{M_{\bullet}}{M_{\star}}\right)^{1/6}, \label{vmaxapp}
\end{equation}
where in the last line we used $\kappa = 0.1$ (note that the dependence on $\kappa$ enters this expression as $\kappa^{1/4}$, i.e., weakly). The time at which this maximum velocity is achieved is
\begin{equation}
    T_{\rm max} \simeq 10\left(\frac{M_{\bullet}}{M_{\star}}\right)^{1/2}. \label{Tmax}
\end{equation}
Figure \ref{fig:vmax} compares this prediction to the result obtained by numerically solving for the $X_0$ at which the line density (Equation \ref{lambdaofT}) is maximized\footnote{We cannot use Equation \eqref{epscapp} because it approximates the difference in the shear by using the value at the marginally bound radius, which is clearly not valid once the peak density shifts into the region of homologous expansion.}, with Equation \eqref{lambda0} for the initial density, for $M_{\bullet}/M_{\star} = 10^2$ (blue), $10^3$ (green), $10^4$ (orange), $10^5$ (purple), and $10^6$ (gray), and we used $\alpha = 3.83$ and $\beta = 0.9$ appropriate to a $5/3$-polytrope; the solid lines are the numerical predictions, the dashed lines result from Equation \eqref{vmaxapp}, and the black-dot-dashed curve is the prediction from Equation \eqref{vkickapp}. Equation \eqref{vmaxapp} provides a good approximation for the asymptotic value approached by the numerical solution. However, the transition to the asymptotic value is not abrupt at the time predicted by Equation \eqref{Tmax} (shown by vertical, dot-dashed lines), but rather takes place over another $\sim (M_{\bullet}/M_{\star})^{1/2}$ dynamical times; this is not surprising, given that the shear smoothly varies from the marginally bound radius to the unbound tail. Nonetheless, the time predicted in Equation \eqref{Tmax} demarcates the point where two solutions with different $M_{\bullet}$ clearly bifurcate from one another. 

The implication of Equation \eqref{vkickapp} is that there is a maximum speed attainable by the core in this model, and it is above the escape speed of the original star when $(M_{\bullet}/M_{\star})^{1/6} \simeq (0.35)^{-1}$, or $M_{\bullet}/M_{\star} \simeq 540$. However, Equation \eqref{Tmax} illustrates that the time at which this maximum speed is approached is many dynamical times of the star, and hence for large black hole masses the kick speed is effectively independent of the SMBH mass (as implied by Equations \ref{vkickfun} and \ref{vkickfun2} and as seen at early times in Figure \ref{fig:vmax}). If, for example, we impose that $T_{\rm max} \lesssim few\times 100$, then the black hole mass dependence of the kick speed is apparent for $M_{\bullet}/M_{\star} \lesssim 10^3$ and for core masses -- from the preceding section -- satisfying
\begin{equation}
    \frac{M_{\rm c}}{M_{\star}} \lesssim 1.7\left(\frac{M_{\bullet}}{M_{\star}}\right)^{-1/2}. \label{McofMbullet}
\end{equation}

\section{Summary and Conclusions}
\label{sec:summary}
Some partially disrupted stars are ejected on positive-energy orbits following their tidal interaction with an SMBH. Here we suggest that the dominant mechanism for generating this effect is not asymmetric mass loss, but rather arises from a phase of quasi-ballistic expansion of the debris post-pericenter and the corresponding differential shear experienced by the material; this effect occurs independently of any potential mass asymmetries. The motivation for this argument is that partial disruptions near the complete-disruption threshold unambiguously lead to cores that reform significantly after pericenter (e.g., \citealt{guillochon13} note that their $\beta = 0.75 - 0.85$ disruptions of 5/3-polytropes resulted in core reformation after $4\times 10^{4}$ s, i.e., at least $\sim 20$ dynamical times post-pericenter; the same statement applies to their $\beta = 1.7-1.8$ disruptions of $4/3$-polytropes). Among the best pieces of evidence to suggest the viability of this interpretation is the $\beta = 0.9$ disruption performed in \citet{gafton15}: the left panel of their Figure 4 demonstrates that the velocity of their ``core'' increases monotonically in the days after disruption; this suggests that their energy condition for singling out the core properties is actually tracing the evolution of the maximum density in their core-less stream. Note that \citet{gafton15} use the same energy condition as employed in \citet{guillochon13} and \citet{manukian13}.\footnote{\citet{guillochon13} discuss this physical shortcoming of this approach, i.e., that their definition yields a non-zero core mass in the absence of a core; for another example, see the top panel of Figure 4 in \citet{goicovic19}.} 

For core reformation times $T\lesssim \sqrt{M_{\bullet}/M_{\star}}$, where $T$ is in units of the dynamical time of the star, this model predicts that the kick imparted to the core scales as the escape speed of the star, is independent of the mass of the SMBH, and grows with core reformation time as $\propto T^{1/3}$, or -- in terms of the core mass $M_{\rm c}$ -- $v_{\rm kick} \propto M_{\rm c}^{-1/3}$. The maximum kick speed is therefore not limited to the escape speed of the star, which agrees with the findings of \citet{gafton15}, who found that some such super-escape-speed cores could be generated once the core mass was sufficiently small. However, since the increase in core energy (velocity) is associated with the difference in the shear between the unbound and bound regions of the material, the increase in core energy with reformation time ceases once the location of maximum density is firmly within the unbound tail. The upper limit to the kick speed can then be estimated as $v_{\rm kick} \simeq 0.4 v_{\rm esc}\left(M_{\bullet}/M_{\star}\right)^{1/6}$, which is reached in $T_{\rm max} \sim 10\left(M_{\bullet}/M_{\star}\right)^{1/2}$. For SMBH masses satisfying $M_{\bullet} \gtrsim 10^3 M_{\star}$, this time is sufficiently late (for reasonable core masses) that the core properties are effectively independent of SMBH mass. However, for $M_{\bullet}/M_{\star} \lesssim 10^3$, this model predicts that cores with comparable (and small; see Equation \ref{McofMbullet}) masses will have kick velocities that scale as $M_{\bullet}^{1/2}$. 

In agreement with the predictions of this model, a positive correlation between kick speed and black hole mass was recovered by \citet{kiroglu23} for black holes with mass $\lesssim 10^3 M_{\odot}$ (see the middle panel of their Figure 4), and those authors attributed the black hole mass dependence to the fact that larger ejection speeds of the tidally stripped material are attained as the black hole mass increases. However, since the ejection speed of the stripped debris scales as $\propto (M_{\bullet}/M_{\star})^{1/6}$ \citep{rees88}, one would expect the correlation with black hole mass to continue into the supermassive regime. Indeed, one could interpret the lack of correlation with supermassive black hole mass as evidence against an asymmetric mass loss origin, given that -- and in agreement with \citet{kiroglu23} -- the larger ejecta speeds should lead to a higher kick speed under this interpretation.

The model presented here is Newtonian and independent of SMBH mass when the core reforms sufficiently rapidly (see Equation \ref{epsilonc}), but the same statement would not necessarily apply when general relativity is incorporated, and the results in \citet{gafton15} show that there are substantial differences when the relativistic gravitational field of the black hole is included. One would predict that the same pericenter distance leads to a larger kick velocity, owing to the fact that the tidal shear of the black hole is larger when relativistic effects are included, and this is the trend found by \citet{gafton15}. Extending the model described in Section \ref{sec:physical} to incorporate the Kerr metric would be straightforward, which would be necessary for accurately constraining the kick velocity when the SMBH mass is large ($\gtrsim 10^{7} M_{\odot}$).

Finally, and in agreement with \citet{manukian13}, the velocity kicks for low-mass stars generated via this process are likely too weak to be consistent with observations of hypervelocity stars (HVSs), many of which possess velocities in excess of $\sim 400$ km s$^{-1}$ at $50-100$ kpc (e.g., \citealt{brown15}). However, high-mass stars -- with initial masses $\gtrsim 20 M_{\odot}$ and significantly larger escape speeds than low-mass stars -- that lose a substantial fraction of their mass could be ejected with initial speeds in excess of $1-2\times 10^3$ km s$^{-1}$, as super-escape speeds are possible (e.g., Figure \ref{fig:vkick}). While their number is likely small, we conclude that this mechanism can produce stars that escape our Galaxy.

\section*{Acknowledgements}
E.R.C.~acknowledges support from NASA through the Astrophysics Theory Program, grant 80NSSC24K0897. C.J.N.~acknowledges support from the Science and Technology Facilities Council (grant No. ST/Y000544/1) and from the Leverhulme Trust (grant No. RPG-2021-380). We thank the anonymous referee for useful feedback that improved the clarity of the work.

\section*{Data Availability} 
The data underlying this article will be shared on reasonable request to the corresponding author.

\bibliographystyle{mnras}

\bsp	
\label{lastpage}
\end{document}